\documentclass[aps,prb,notitlepage,twocolumn]{revtex4-2}

\usepackage{amsmath}
\usepackage{amssymb}
\usepackage{graphicx}
\usepackage{epstopdf}
\usepackage[colorlinks=true]{hyperref}
\usepackage{xcolor}
\usepackage{scalerel}
\usepackage[utf8]{inputenc}
\usepackage{amsfonts}
\usepackage{graphicx}
\usepackage{dcolumn}
\usepackage{bm}
\usepackage{hyperref}
\usepackage{accents}

\usepackage[cjk]{kotex}
\CJKscale{0.9}

\newcommand{\bd}[1]{\mathbf{#1}}
\newcommand{\pmat}[1]{\begin{pmatrix} #1 \end{pmatrix}}

\newcommand{\br}[1]{[ \, #1 \, ]}
 \newcommand{\temp}[1]{T_{ \mathrm{ #1 } } }

\begin{document}
\title{Interaction of gapless spin waves and a domain wall in an easy-cone ferromagnet}

\author{Wooyon Kim}
\affiliation{Department of Physics, Korea Advanced Institute of Science and Technology, Daejeon 34141, Republic of Korea}
\author{Se Kwon Kim}
\affiliation{Department of Physics, Korea Advanced Institute of Science and Technology, Daejeon 34141, Republic of Korea}

\begin{abstract}
We theoretically study the interaction of spin waves and a domain wall in a quasi-one-dimensional easy-cone ferromagnet. The gapless spin waves on top of a domain wall are found to exhibit finite reflection in contrast to the well-known perfect transmission of gapful spin waves in easy-axis magnets. Based on the obtained scattering properties, we study the thermal-magnon-driven dynamics of a domain wall subjected to a thermal bias within the Landauer-Büttiker formalism, where transmitted magnons are shown to exert the magnonic torque on the domain wall and thereby drive it with the velocity linear to the applied thermal bias. The peculiar gapless nature of spin waves in easy-cone magnets enables the thermally-driven domain-wall motion even at low temperatures, differing from the easy-axis case where the domain-wall velocity is exponentially suppressed at low temperatures. Our work suggests that easy-cone magnets can serve as a useful platform to study the interaction of gapless spin waves and nonlinear excitations and thereby realize low-temperature magnon-related phenomena.
\end{abstract}

\maketitle 

\section{introduction}
Magnetic systems can hold a variety of topological defects, such as domain walls (DWs), vortices and skyrmions, and their dynamics have been studied for decades both for fundamental and practical interest~\cite{solitons}
. In particular, magnetic DWs have drawn great attention due to their practical applications exemplified by domain-wall race track memory~\cite{racetrack}. The mechanism of driving a DW has been studied extensively in several easy-axis magnets with various driving means, such as an external magnetic field~\cite{H-field1, H-field2}, and an electric current~\cite{electric_stt1_berger,electric_stt2_slonczewski,stt1,nstt1,nstt2,sot,nsot1}. In addition to these means, spin waves also have been shown to be able to drive a DW~\cite{sw1_shen, sw2,nsw1,nsw2}. Magnons, quanta of spin waves, carry spin $\hbar$ in the opposite direction to the background spin, which enables them to push a DW by flipping their spin while moving across a  DW and thereby transfer spin angular momentum to the DW according to the conservation of the total angular momentum. For easy-axis ferromagnets and antiferromagnets, it is well known that spin waves show perfect transmission through a DW~\cite{magnonic_STT_JoovonKIM,magnonic_stt_FM,magnonic_stt_AFM_tveten,magnonic_stt_AFM_SKK}. Magnons in the easy-axis magnets have a finite gap to be excited and therefore the effects of thermal magnons on transport properties are exponentially suppressed at sufficiently low temperatures compared to the magnon gap. 

\begin{figure}
    \centering
    \includegraphics[width=\columnwidth]{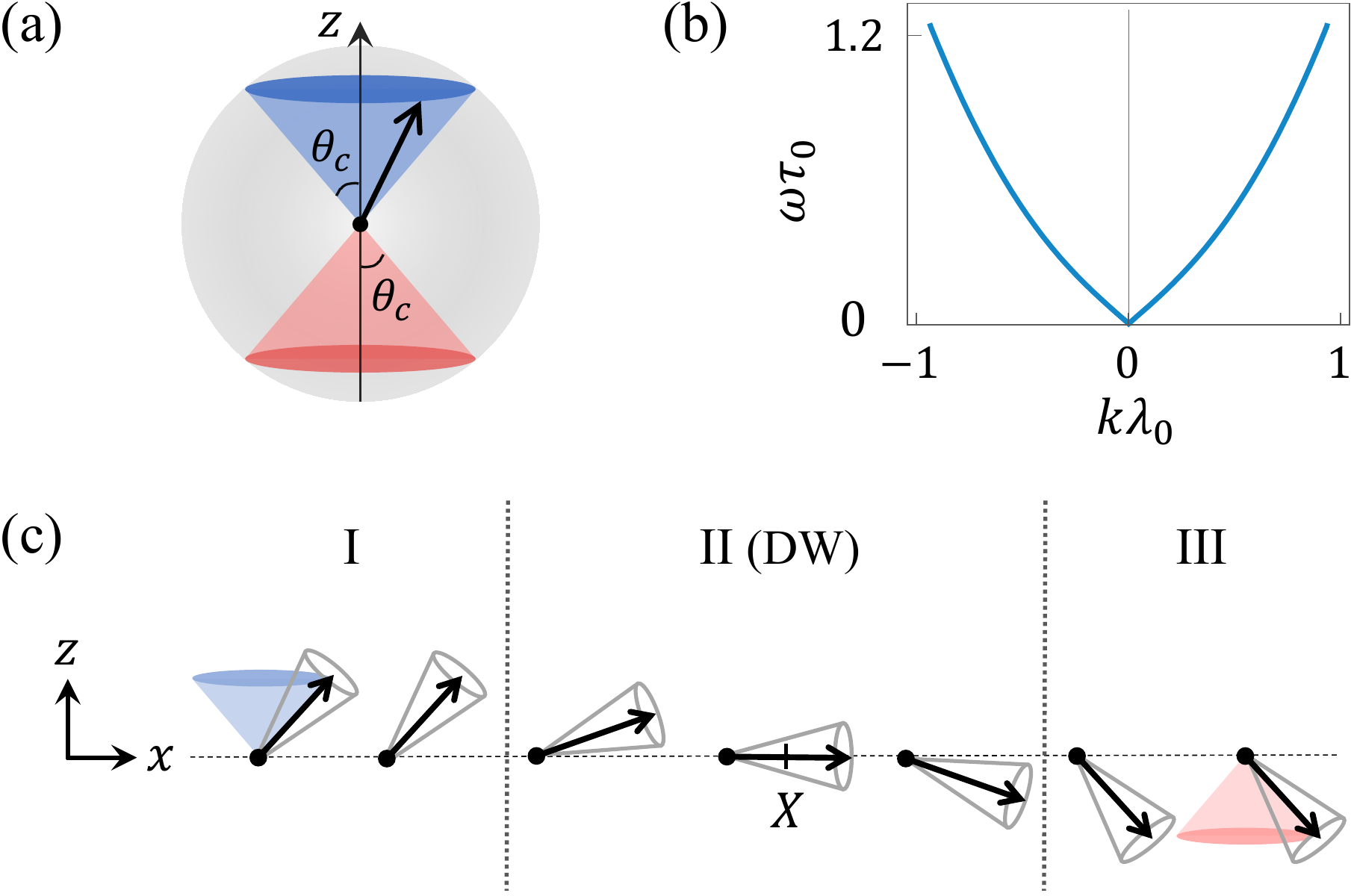}
    \caption{(a) Ground-state manifolds of an easy-cone magnet in the unit sphere. The manifolds form a pair of cones about the z-axis with angle $\theta_{\mathrm c}$ (upper blue) and $\pi-\theta_{\mathrm c}$ (lower red). The black arrow on the blue cone represents one possible ground-state magnetization. (b) The gapless dispersion relation $\omega ( k ) $ of spin waves in a ground state of an easy-cone magnet. Both axes are in natural units of length and time, $\lambda_0$ and $\tau_0$. See the main text for their definitions. (c) Schematic illustration of a one-dimensional easy-cone magnet with a domain wall (DW). Region I(III) is in the ground state with uniform magnetization within the blue(red) cone. These regions are interpolated by the region II, where the DW is located (centered at $X$). The black arrows represent the magnetizations in a static DW profile, while the gray cones around them represent time-evolving magnetizations of spin waves.}
    \label{fig:cone}
\end{figure}

Easy-cone magnets are another class of magnets having uniaxial anisotropy, in which ground-state manifolds form a  couple of cones about the high-symmetry axis \cite{ncone1,ncone2,ncone3,ncone4} [See Fig.~\ref{fig:cone}(a)]. The ground states spontaneously break both the $\mathrm{Z_2}$ and U(1) symmetries by choosing, respectively, which of the cones to reside in and which direction within the cone is pointed at. The spontaneous breaking of these symmetries endows easy-cone magnets with the peculiar ability to support DWs [Fig.~\ref{fig:cone}(c)] and gapless spin-wave excitations [Fig.~\ref{fig:cone}(c)] in one system~\cite{symmetry_breaking}, which is absent in easy-axis magnets. In particular, the latter gapless property of spin waves in these magnets is expected to enable us to drive a DW with spin waves of lower energy compared to the case of a gapful easy-axis magnet. Recent technology to fabricate easy-cone magnets has advanced significantly, increasing the feasibility of easy-cone magnets for experimental studies and technological applications. The example materials include Co/Pt, Ta/Co$_{60}$Fe$_{20}$B$_{20}$/MgO, (Cr$_{0.9}$B$_{0.1}$)Te, and NdCo$_5$~\cite{CoPt,CoFeB,CrBTe,ndco5_1,ndco5_2,ndco5_3,ndco5_4}. However, despite the aforementioned unique features of easy-cone systems compared to easy-axis systems, the spin-wave properties, the dynamics of solitons such as a DW, and their interactions in easy-cone magnets have been studied little. 

 In this paper, we study the interaction of spin waves and a DW in an easy-cone ferromagnet within the Landau-Lifshitz framework. In Sec.~\ref{sec.cone}, we begin by giving a brief review of the easy-cone system and the DW solution therein. In Sec.~\ref{sec.sw}, we develop a theory for spin waves on top of a DW and compute the reflection and the transmission probability of the spin waves interacting with the DW. Using the scattering property obtained, in Sec.~\ref{sec.thermal motion}, we study the magnon-driven DW motion subjected to a thermal bias and obtain the DW velocity as a function of the temperatures involved, within the Landauer-B\"uttiker formalism. In particular, we show that the gapless nature of spin waves in easy-cone magnets gives rise to the finite DW velocity even at low temperatures, in contrast to the easy-axis case where the analogous DW velocity is exponentially suppressed as the temperature decreases due to the finite spin-wave gap.

\section{easy-cone systems}\label{sec.cone}

In this section, we describe easy-cone magnets, their symmetry properties, and ground states. We also present an exact solution for a domain wall.

\subsection{Easy-cone magnets}

We consider a quasi-one-dimensional ferromagnet along the $x$-axis with the potential energy given by 
\begin{equation}
    U[\mathbf m(x)]=\int\Big[\;\frac{A}{2}(\partial_x \mathbf{m})^2-\frac{K}{2}m_z^2 + \frac{K'}{2}m_z^4\;\Big ] dx\,,
\end{equation}
where $A > 0$ is the exchange coefficient, $ K > 0 $ is the effective first-order uniaxial anisotropy, and $K' > 0$ is the second-order uniaxial anisotropy \cite{ncone1,ncone2,ncone3,ncone4}. Here, $K = K_{\text u} - \mu_0 M_{\text s} ^{ 2} / 2$ includes the dipolar-induced shape anisotropy as well as the magnetocrystalline anisotropy $K_u$.  Note that the first-order anisotropy favors the magnetization along the $z$-axis, while the second-order anisotropy tends to tilt the magnetization away from the $z$-axis. To parametrize the competition of these two effects, we define a dimensionless number $\kappa = K / (2K')$. The condition for the system to be an easy-cone magnet is given by $\kappa < 1$ \cite{ncone2}, which will be assumed throughout the paper. The potential energy possesses two distinct symmetries. Firstly, it is invariant under time-reversal $\mathbf{m} \mapsto - \mathbf{m}$, showing the Z$_2$ symmetry of the system. Secondly, it is invariant under global rotations of the magnetization $\mathbf{m}(x) \mapsto \hat{R}_z (\varphi) \mathbf{m}(x)$, where $\hat{R}_z (\varphi)$ is a three-dimensional rotation matrix about the $z$-axis with angle $\varphi$, showing the U(1) spin-rotational symmetry of the system.

Given the potential energy, the couple of cones of the ground-state manifold [Fig.~\ref{fig:cone}(a)] are determined as follows. Their angle about the high-symmetry axis ($z$-axis) are $\theta_{\mathrm c} = \arccos \sqrt{\kappa} \left(< \pi/2 \right) $ for the upper blue cone and $\pi - \theta_{\mathrm c}$  for the lower red cone, respectively. Note that the angles are determined by $\kappa$, the relative strength of the first- and the second-order anisotropies, as expected. A ground state is a uniform array of the magnetization $\bd m$ which belongs to one of the given manifolds. It is well described by the spherical coordinates $\theta$ and $\phi$ where $\bd m = ( \sin \theta \cos \phi , \, \sin \theta \sin \phi , \, \cos \theta )$, and is given by 
\begin{align}
\theta ( x ) &= \theta_{\mathrm c} \quad \text{or} \quad \pi - \theta_{\mathrm c} \, , \\
\phi ( x ) &= \Phi \, ,
\end{align}
in which each coordinate breaks one of two symmetries: $\theta$ breaks Z$_2$ symmetry and $\phi$ breaks U(1) spin-rotational symmetry.

For the following discussions, it is convenient to use the following natural units of length, time and energy:
\begin{equation}
\label{eq:unit}
\lambda_0=\sqrt{A/K}\,,\quad \tau_0=s/K\,,
    \quad \epsilon_0=\sqrt{AK} \, ,
\end{equation}
where $s$ is the spin density.

\subsection{Domain walls}

The easy-cone system supports another stable state referred to as a DW, which connects two uniform ground states in different manifolds while minimizing the potential energy. That is, a DW is a stationary solution satisfying $\delta U / \delta \theta_0 = 0 $ and $\delta U / \delta \phi_0 = 0$, with boundary conditions $\theta_0 ( x \rightarrow \mp \infty ) = \theta_{\mathrm c}$ and $\theta_0 ( x \rightarrow \pm \infty) = \pi - \theta_{\mathrm c} \,$. The exact solution is available \cite{cone2_SKK,cone3_peace_jang} which is given by 
\begin{align}
\theta_0 ( x ) &= \frac{\pi}{2} \pm \arctan \left \{ \sqrt{\frac{\kappa}{1 - \kappa}} \tanh \left [ \sqrt{ \kappa ( 1 - \kappa ) } \,\, \frac{\, x - X \,}{\lambda_\text{d} } \right ] \right \} \label{DW sol1} \\
\phi_0 (x) &\equiv \Phi \label{DW sol2} \, ,
\end{align}
where $\lambda_\text{d} = \sqrt{2 \kappa} \, \lambda_0 $ is the DW width. See Fig.~\ref{fig:cone}(c) for the schematic illustration of the DW described by the Eqs.~\eqref{DW sol1} and \eqref{DW sol2} with plus sign and $\Phi = 0$. Here, the polar angle changes from $\theta_{\mathrm c}$ to $\pi - \theta_{\mathrm c}$ as $x$ varies from $- \infty$ to $\infty$ while the azimuthal angle is uniform. In Eq.~(\ref{DW sol1}), $X$ represents the DW center at which the polar angle is $\pi/2$. Due to the translational invariance of our system, $X$ is arbitrary and thus represents the zero-energy mode associated with spontaneous breaking of the translational invariance by the DW. In this paper, we are interested in spin waves on top of this DW, which we turn to below.

\section{Interaction of spin waves with a domain wall} \label{sec.sw}

In this section, we study spin-wave dynamics on top of a DW. In particular, we  confirm the gapless dispersion and show that the incident spin wave is partially reflected from the DW regardless of its frequency, which is in contrast to the well-known perfect transmission in easy-axis magnets \cite{magnonic_STT_JoovonKIM,magnonic_stt_FM,magnonic_stt_AFM_tveten,magnonic_stt_AFM_SKK}.

\subsection{Spin waves}

In order to consider a spin wave on top of a DW, we divide the magnetization $\bd m (x, t) $ into the static DW profile $\bd m_0 (x \, ; X , \Phi )$ and a small perturbation $\delta \bd m$. In the local frame, the small perturbation can be written as
\begin{equation}
\delta \bd m (x, t) \approx \bd{\hat \theta} \, \delta_1 (x,t) + \bd{\hat \phi} \, \delta_2 (x,t) \, ,
\end{equation}
where $\bd{\hat \theta} = \partial \bd m_0 / \partial \theta$, $\bd{\hat \phi} = ( 1/ \sin \theta_0 ) ( \partial \bd m_0 / \partial \phi ) , $ and $\bd m_0$ form the local orthonormal frame.

\begin{figure}
    \centering
    \includegraphics[width=0.96\columnwidth]{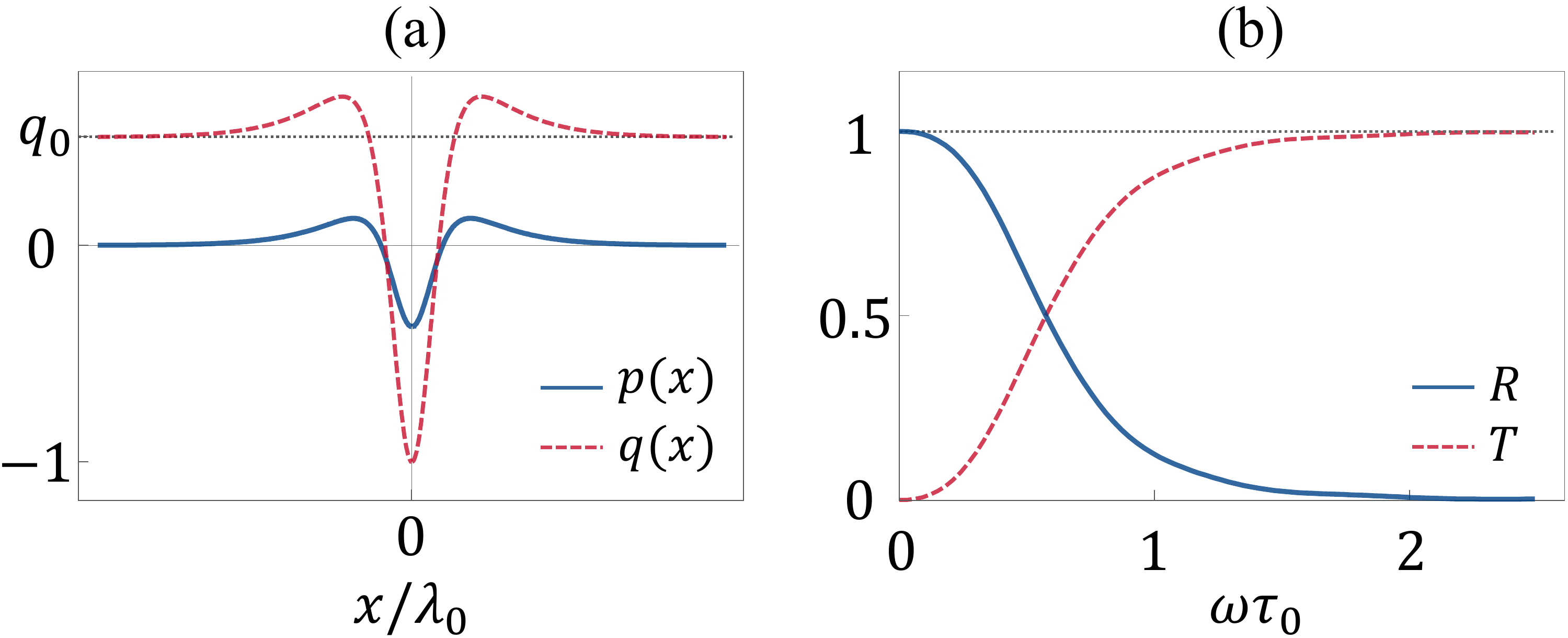}
    \caption{(a) The plots of $p(x)$ (blue solid) and $q(x)$ (red dashed line) in the spin-wave equation~\eqref{pq} with $q_0 = 1 $. Both are even functions and converge to constant values far away from the DW. The middle region where $p(x)$ and $q(x)$ vary corresponds to the DW. (b) The probability of reflection $R$ (blue solid) and transmission $T$ (red dashed) of a spin wave scattering with a DW, as a function of frequency $\omega \tau_0$. Their sum $R + T$ shown as the dotted black line is confirmed to be unity. Both (a) and (b) are for $\theta_{\mathrm c} = \pi / 4 \, .$}
    \label{fig:sw}
\end{figure}

By linearizing the Landau-Lifshitz equation~\cite{llg_eq} about the DW solution while neglecting the damping
\begin{equation}
\frac{\partial \bd m}{\partial t} = -\frac{\delta U}{\delta \bd m} \times \bd m \, ,
\end{equation}
in the local coordinate system, we have a set of first-order equations in the small field $\delta \bd m$ given by
\begin{subequations}
\begin{align}
- \dot \delta_1 (x, t) &= -\delta_2 '' + p(x) \delta_2 \, , \\
\dot \delta_2 (x, t) &= -\delta_1 '' + q(x) \delta_1\, ,
\end{align}\label{swDW}\,
\end{subequations}
with
\begin{subequations}
\begin{align}
p(x) &= - ( 2 \kappa ) ^{-1} \left[ \, 3 \cos ^4 \theta_0 (x) - 4 \kappa \cos ^2 \theta_0 (x) + \kappa ^2 \, \right] \, , \\
q(x) &= - \kappa ^{-1} \left[ \, 4 \cos ^4 \theta_0 (x) - ( 2 \kappa + 3 ) \cos ^2 \theta_0 (x) + \kappa \, \right] \label{pq} \, . 
\end{align} 
\end{subequations}
See Fig.~\ref{fig:sw}(a) for the exemplary plots of $p(x)$ and $q(x)$ for $\theta_{\mathrm c} = \pi / 4$. Note that $p(x)$ approaches zero as $x \rightarrow \pm \infty$, which stems from the spontaneous breaking of the U(1) spin-rotational symmetry in ground states and is associated with the gapless nature of spin waves therein. However, $q(x)$ approaches a finite value $q_0 = 2 ( 1 - \kappa ) $ as $x \rightarrow \pm \infty$, representing the finite energy cost for the magnetization to be tilted away from the easy-cone manifolds. 

Far away from the DW where $p(x)$ and  $q(x)$ are uniform, the spin-wave equation has plane-wave solutions $\delta_n (x,t) = A_n \exp (i k x - i \omega t)$ with the dispersion relation given by
\begin{equation}
\omega (k ) = \sqrt{ k^2 ( k^2 + q_0 )} \, .
\end{equation}
See Fig.~\ref{fig:cone}(b) for the plot and note that the gap is zero.

\subsection{Reflection and transmission probability}
The incident spin wave is found to be partially reflected from the DW. The probability of reflection and transmission were obtained by numerically solving Eqs.~\eqref{swDW} within the Green's function formalism detailed in the Appendix~\ref{app}. See Fig.~\ref{fig:sw}(b) for the plots of the probabilities for $\theta_{\mathrm c} =  \pi / 4$, which represents one of our main results. Note the finite reflection probability at the whole energy ranges, which is in contrast to the reflectionless spin waves in the easy-axis counterpart. The transmission probability increases as the spin-wave energy increases, since the effect of the energy barrier on its transmission becomes weaker at high energies.

When a spin wave is quantized, a quasiparticle referred to as a magnon emerges. The reflection probability and the transmission probability of magnons are the same as the ones for the spin waves, which we will invoke below when discussing the thermal-magnon-driven DW motion.

 \begin{figure}
    \centering
    \includegraphics[width=\columnwidth]{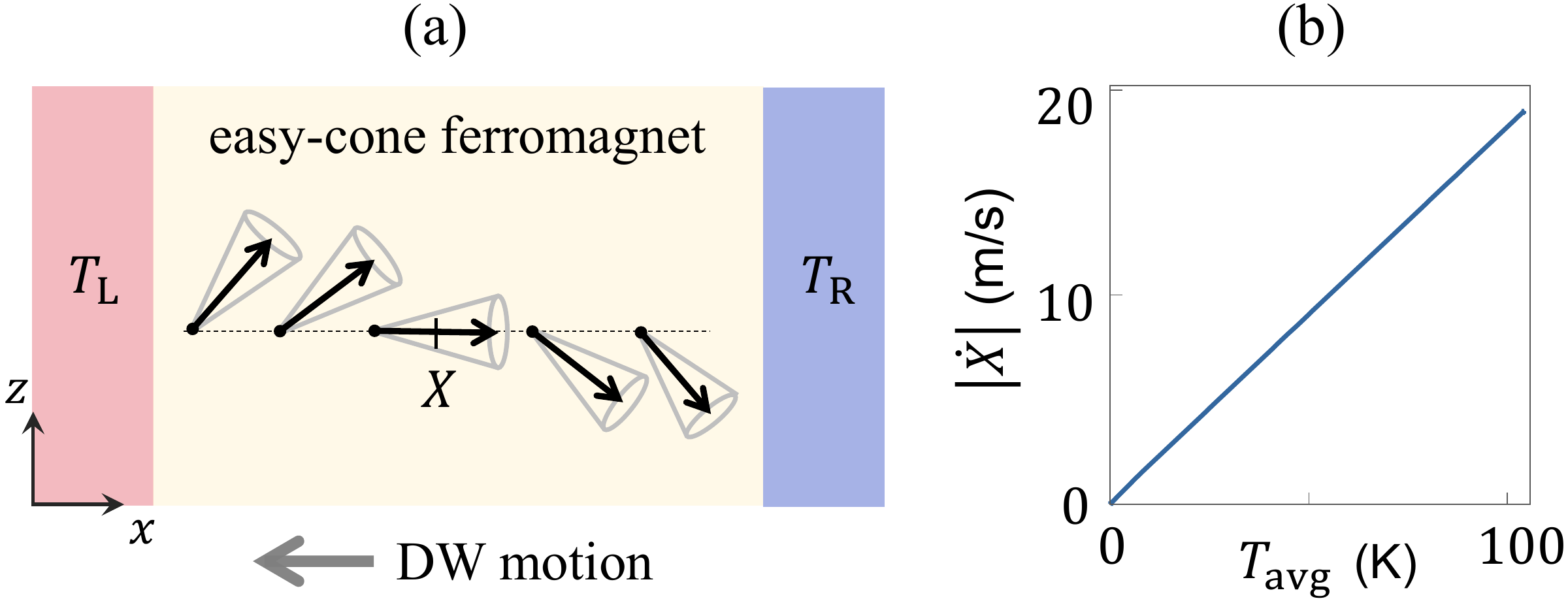}
    \caption{(a) A schematic illustration of a setup for a thermally driven DW. An easy-cone ferromagnet with a DW is placed between two thermal reservoirs that maintain constant temperatures $T_L$ and $T_R  (< T_L)$, respectively. Each reservoir injects thermal magnons into the magnet, which exert a torque on the DW and thereby push it to the region with higher temperature. (b) DW speed in the setup (a), as a function of $\temp{avg} = ( \temp L + \temp R ) / 2 $. The temperature difference $\Delta T = \temp L - \temp R$ is set to $0.1 \temp{avg}$. For the used material parameters, see the main text.}
\label{fig:DW motion}
\end{figure}

\section{Domain-Wall motion driven by thermal magnons} \label{sec.thermal motion}

In this section, we study the DW motion driven by the scattering of thermal magnons. See Fig.~\ref{fig:DW motion}(a) for the schematic illustration of a system. An easy-cone magnet with a single DW is placed between the left and the right thermal reservoirs that maintain finite temperatures $\temp{L}$ and $\temp{R}$, respectively. We employ the Landauer-B\"uttiker formalism by assuming ballistic magnon transport to obtain the DW motion~\cite{landauer_landauer-buttiker,buttiker_landauer-buttiker,Datta_landauer-buttiker}, as done in Refs.~\cite{meier,DW_momenta1}. In Eq.~\eqref{Vdw}, we present the DW velocity as a function of the average temperature $\temp{avg} = ( \temp L + \temp R ) / 2$, which is another main result of ours.

\subsection{Magnonic force and torque}

Thermally populated magnons move between the two thermal reservoirs. When traveling magnons are transmitted through a DW, they exert a torque on it by changing their spin. The torque by the right-moving magnons, which come out of the left reservoir, is given by
\begin{equation}
\tau_{\mathrm{L}} =  \int d \epsilon \,\,\, 2 \hbar \cos \theta_{\mathrm c} \cdot T ( \epsilon ) \cdot n_{\text B} \bigg( \frac{\epsilon}{k_\mathrm{B} \temp L } \bigg) \frac{1}{2 \pi \hbar} \, ,
\end{equation}
with polarization along the $z$-direction, where $n_{\text B}$ is Bose-Einstein distribution function and $k_{\text B}$ is Boltzmann constant. Here, the factor $2 \hbar \cos{\theta_{\mathrm c}} = \hbar \cos{\theta ( x \rightarrow - \infty ) } - \hbar \cos{\theta (x \rightarrow \infty )} $ represents the angular-momentum transfer from a single magnon to the DW, $T(\epsilon)$ is the transmission probability of magnons with energy $\epsilon$, and the last factor $1 / ( 2 \pi \hbar )$ comes from the product of a density of states per unit length and the group velocity $d \omega / dk$ \cite{meier}. Similarly, the torque $\tau_{\mathrm{R}}$ exerted on the DW by the magnons moving from the right reservoir to the left reservoir can be obtained. The net torque $\tau = \tau_{\mathrm L} + \tau_{\mathrm R }$ can be approximated by
\begin{equation}
\tau \approx \frac{\cos \theta_{\mathrm c}}{\pi} \Delta T \int d \epsilon \,\,\, T ( \epsilon ) \cdot \frac{\partial n_{\text B}}{\partial T} \bigg| _{\temp{avg}} \label{torque} \, ,
\end{equation}
for $\Delta T \ll \temp{avg}$, where $\Delta T = \temp L - \temp R$ is the temperature difference between the two reservoirs.

A reflected magnon, however, exerts a force on the DW. By the derivation analogous to the above torque case, the net force exerted on the DW by magnons coming out of the two reservoirs can be approximated by 
\begin{equation}
F \approx \frac{k}{\pi} \Delta T \int d \epsilon \,\,\, R (\epsilon) \cdot \frac{\partial n_{\text B}}{\partial T} \bigg| _{\temp{avg}} \, ,
\end{equation}
where $k / \pi$ is a product of $2 \hbar k$---the linear-momentum transfer by a single magnon--- and $1 / (2 \pi \hbar )$---the product of a density of states per unit length and the group velocity. Here, $R (\epsilon)$ is the reflection probability of magnons with energy $\epsilon$ by the scattering with the DW. Again, the approximation is valid for $\Delta T \ll \temp{avg} \, $.

\subsection{DW motion}

The torque and the force on the DW by the scattering with thermal magnons give rise to the DW motion as follows.
The equations of motion for the DW parameters, the position $X$ and the angle $\Phi$, are given by (in natural units defined in Eq.~(\ref{eq:unit}))
\begin{subequations}
\begin{align}
g \dot X &= \tau \, , \\
-g \, \dot \Phi &= F \, ,
\end{align}
\end{subequations}
where $g = - 2 \cos \theta_{\mathrm c}$ is the gyrotropic coupling constant between $\dot X$ and $\dot \Phi$ ~\cite{thiele}. For the simplicity, we neglect the effects of the damping in the equations of motion. Here, the left-hand sides are the time derivatives of the spin angular momentum and the conserved linear momentum of a DW, respectively, which are derived through the Noether's theorem~\cite{DW_momenta1,DW_momenta2}. The former $ g \dot X$ can be easily understood by considering the dependence of the total spin on the DW position, which determines the lengths of the spin-up and spin-down regions: $S^z _{\text{tot}} \propto \pm X$ (Minus sign for increasing $S^z (x) \, ,$ which is the case of Fig.\ref{fig:DW motion}(a)). For the torque in Eq.~\eqref{torque}, the DW velocity is given by 
\begin{equation}
\dot X = - \frac{\Delta T}{2 \pi} \int d \epsilon \,\,\, T (\epsilon) \cdot \frac{\partial n_{\text B}}{\partial T} \bigg| _{\temp{avg}} \label{Vdw}
\end{equation}
in natural unit [Eq.~(\ref{eq:unit})]. We numerically obtained the velocity by using the material parameters of $\text{NdCo}_5$ given by lattice constant $a = 0.5 \, \mathrm{nm}, \,$ saturation magnetization $M_{\mathrm s} = 1.1 \times 10^6 \, \mathrm{A/m},  A = 1.1\times 10^{-11} \, \mathrm{J/m}, \, K = 2.4 \times 10^6 \, \mathrm{J/m^3}, \, K' = 1.6 \times 10^6 \, \mathrm{J/m^3}$, and thereby $\kappa = 0.75$~\cite{ndco5_1,ndco5_2,ndco5_3,ndco5_4}. Figure ~\ref{fig:DW motion}(b) shows the velocity of the thermally driven DWs in $\mathrm{NdCo_5}$ as a function of $T_{\text{avg}}$ with $\Delta T = 0.1 T_{\text{avg}}$. Note that there is no exponential suppression of the DW velocity as the average temperature decreases, which can be attributed to the gapless nature of spin waves in easy-cone magnets.

\section{summary}
We have studied how spin waves and a DW interact in a one-dimensional easy-cone ferromagnet within the Landau-Lifshitz phenomenology~\cite{llg_eq}. Specifically, we have studied the scattering properties of spin waves with a DW and obtained the reflection and the transmission probability as a function of the wave frequency. Based on this, we have further investigated the magnon-current-driven dynamics of a DW under a thermal bias within the Landauer-Büttiker formalism~\cite{landauer_landauer-buttiker,buttiker_landauer-buttiker,Datta_landauer-buttiker}. The DW velocity is shown to be linear to the applied thermal bias and to increase as an average temperature rises. In particular, the gapless feature of magnons is shown to enable the thermal DW motion at low temperatures without exponential suppression.

In the future, it might be worth investigating a two-dimensional easy-cone magnet that harbors a one-dimensional DW with chiral spin rotation along with it. Here, one can expect magnon deflection by the emergent magnetic field, which is formed exclusively at the DW~\cite{emergent_1,emergent_2,emergent_3,emergent_4}. The analogous research has been conducted for easy-axis ferromagnets with gapful magnons in Ref.~\cite{Chiral_DW_skk}. Compared to the ferromagnetic case, we expect that the easy-cone counterpart would allow us to study the interaction of gapless magnons and a DW with chiral spin rotation.

\begin{acknowledgements}
We thank Ehsan Faridi, Gyungchoon Go, and Giovanni Vignale for the useful discussion. This work was supported by Brain Pool Plus Program through the National Research Foundation of Korea funded by the Ministry of Science and ICT (NRF-2020H1D3A2A03099291), National Research Foundation of Korea(NRF) grant funded by the Korea government(MSIT) (NRF-2021R1C1C1006273), National Research Foundation of Korea funded by the Korea Government via the SRC Center for Quantum Coherence in Condensed Matter (NRF-2016R1A5A1008184), and Basic Science Research Program through the KAIST Basic Science 4.0 Priority Research Center funded by the Ministry of Science and ICT.
\end{acknowledgements}

\appendix

\section{Scattering of spin waves with a domain wall}
\label{app}
We present a numerical method to compute the scattering parameters of spin waves. To compute them, it is important to include evanescent waves $(\sim e^{\pm \kappa x})$ in addition to a plane-wave solution~\cite{evanescent_waves}. Here $\kappa$ is an imaginary wave number with dispersion relation $\omega (\kappa ) = \sqrt{ \kappa ^2 ( \kappa ^2 - q_0 ) }\, .$ (Do not be confused with $\kappa = K / (2K')$ in the main text.) We set an ansatz
\begin{align}
    \Psi_i &= \pmat{ e^{ik x_i} + r e^{-ik x_i} \\ r' e^{\kappa x_i} } \, , \, \text{for } i = 0, -1, -2, \cdots \, , \\
    \Psi_i & = \pmat{ t e^{ik x_i} \\ t' e^{-\kappa (x_i - x_N) } } \quad , \, \text{for } i = N+1, N+2, \cdots \, ,
\end{align}
 outside the DW, which lies in the $i = 1, \, 2, \, \cdots , \, N$ th sites. Here $e^{ikx}$ represents an incoming spin wave, $r \, (r') \, \text{ and } t \, (t')$ is the reflection and transmission coefficient, respectively for the travelling (evanescent) mode.

The recurrence relation of the spinors, $\Psi_i$ is in the form of 
\begin{align}
\omega \Psi_i &= -\br{t} \Psi_{i-1} + \br{s_i} \Psi_i - \br{t} \Psi_{i+1} \, . \label{recurrence}
\end{align}
Here the $\br t$ and $\br{s_i}$ are $2 \times 2$ matrices, which can be specified from the spin-wave equation \eqref{swDW}.

Note that the relations for the spinors at the DW boundaries---i.e., the ones for (-1, 0, 1)-th or $( N, \, N + 1, \, N + 2 )$-th spinors---will give equations for $r \, ,  r' \, , t \, , t'$ , once we know $\Psi_1$ and $ \Psi_N$ that are inside the DW. To this end, let us define the propagator $\br G$, which is the $N \times N$ matrix whose elements are $2 \times 2$ matrices :
\begin{equation}
   \br{G}_{ij} = \begin{cases}
        \quad \br{t} & \text{ , } j = i \pm 1 \\
        \quad \br{ \omega I - s_i } & \text{ , } j = i  \\ 
        \quad \br 0 & \text{ , otherwise }
    \end{cases} \,\, ,
\end{equation}
where $\br I $ is the $2 \times 2$ identity matrix. The whole spinors inside the DW including $\Psi_1 \text{ and } \Psi_N$ are determined : 
\begin{equation}
    \pmat{ \Psi_1 \\ \Psi_2 \\ \vdots \\ \Psi_{N-1} \\ \Psi_N } = \br G \pmat{ -\br{t} \Psi_0 \\ 0 \\ \vdots \\ 0 \\ -\br{t} \Psi_{N+1} } \, .
\end{equation}

Therefore, Eq.\eqref{recurrence} of $i=0$ and $N+1$ reduce to a set of four linear equations for $r, \, t, \, r', \, t'$ :
\begin{align}
    -\br{t} \Psi_{-1} + \br{-\omega I + s_0 + t \, G_{11} t} \Psi_0 + \br{t \, G_{1N} t} \Psi_{N+1} &= \mathbf 0 \, , \\
    \br{t \, G_{N1} t} \Psi_0 + \br{-\omega I + s_{N+1} + t \, G_{NN} t} \Psi_{N+1}
 - \br{t} \Psi_{N+2} &= \mathbf 0 \, ,
 \end{align}
from which we obtain the probability of reflection, $|r|^2$, and of transmission, $|t|^2$.

\bibliographystyle{apsrev4-2}
\bibliography{bib.bib}

\end{document}